\documentclass[pre,
 amsmath,amssymb,twocolumn,
 aps,superscriptaddress
]{revtex4-2}

\usepackage[toc,page]{appendix}
\usepackage[normalem]{ulem}
\usepackage[dvipsnames]{xcolor}
\usepackage{graphicx}
\usepackage{dcolumn}
\usepackage{bm}
\usepackage{float}
\usepackage{colortbl}

\usepackage{braket}

\newcommand{\beq}{\begin{equation}}
\newcommand{\eeq}{\end{equation}}

\begin{document}
\preprint{APS/123-QED}

\title{New algorithm to determine a minimal representation of the molecular surface}

\author{Greta Grassmann}

\affiliation{Universit\`a di Bologna, Via Irnerio 46, Bologna I-40126, Italy.}

\author{Mattia Miotto} 
\affiliation{Center for Life Nanoscience-\& Neuro-Science, Fondazione Istituto Italiano di Tecnologia, Viale Regina Elena 291, 00161 Rome, Italy.}

\author{Lorenzo Di Rienzo}
\affiliation{Center for Life Nanoscience-\& Neuro-Science, Fondazione Istituto Italiano di Tecnologia, Viale Regina Elena 291, 00161 Rome, Italy.}

\author{Giorgio Gosti}
\affiliation{Center for Life Nanoscience-\& Neuro-Science, Fondazione Istituto Italiano di Tecnologia, Viale Regina Elena 291, 00161 Rome, Italy.}

\author{Giancarlo Ruocco}
\affiliation{Center for Life Nanoscience-\& Neuro-Science, Fondazione Istituto Italiano di Tecnologia, Viale Regina Elena 291, 00161 Rome, Italy.}
\affiliation{Department of Physics, Sapienza University, Piazzale Aldo Moro 5, 00185 Rome, Italy.}

\author{Edoardo Milanetti}
\affiliation{Department of Physics, Sapienza University, Piazzale Aldo Moro 5, 00185 Rome, Italy.}

\affiliation{Center for Life Nanoscience-\& Neuro-Science, Fondazione Istituto Italiano di Tecnologia, Viale Regina Elena 291, 00161 Rome, Italy.}


\begin{abstract}
Most proteins perform their biological function by interacting with one or more molecular partners. In this respect, characterizing the features of the molecular surface, especially in the portions where the interaction takes place, turned out to be a crucial step in the investigation of the mechanisms of recognition and binding between molecules. Predictive methods often rely on extensive samplings of molecular patches with the aim to identify hot spots on the surface. 
In this framework, analysis of large proteins and/or many molecular dynamics frames is often unfeasible due to the high computational cost. Thus, finding optimal ways to reduce the number of points to be sampled maintaining the biological information carried by the molecular surface is pivotal. 
Here, we present a new theoretical and computational algorithm with the aim of determining a subset of surface points, appropriately selected in space, in order to maximize the information of the overall shape of the molecule by minimizing the number of total points.
We test our procedure by looking at the local shape of the surface through a recently developed method based on the formalism of \textit{Zernike} polynomials in two dimensions, which is able to characterize the local shape properties of portions of molecular surfaces. The results of this method show that a remarkably higher ability of this algorithm to reproduce the information of the complete molecular surface compared to uniform random sampling.
\newline
\newline
\emph{keywords:} \textit{Zernike} polynomial expansion, Molecular surface, Protein interaction, Optimization method.
\end{abstract}
\maketitle

\section{Introduction}

Interactions between biomolecules play a fundamental role for most of cellular processes, from DNA replication to protein degradation \cite{keskin2008principles,nooren2003structural,perkins2010transient}. 
Since it has been estimated that over $80\%$ of proteins operate in molecular complexes \cite{berggaard2007methods}, many computational approaches has been developed for investigating and/or predicting protein–protein interactions (PPIs) \cite{vangone2015contacts,qin2011automated,audie2007novel}.

From this point of view, the definition of the molecular surface plays a fundamental role \cite{nicolau2013protein, Moreira2015,Xue2015,Vakser2014,deVries2008,Brender2015}. Indeed, at short distances, the shape complementarity between interacting regions dictates the stabilizing role exerted by van der Waals interactions and: thus, the shape of local surface regions has a key role in predicting protein ability to bind its molecular partner \cite{Teyra2010}. Many algorithms have been designed in order to build the molecular surface at a resolution scale similar to the size of the protein molecular partner\cite{can2006efficient}, for example up to 5 \AA, which is approximately the size of a large solvent molecule. The variation in the size of the probe molecule affects the level of detail of the molecular surface \cite{nicolau2013protein} as much as the density of points used in the surface representation.

Computational methods that aim at the identification of hot-spots and/or binding regions often perform extensive samplings of molecular surface portions, or patches~\cite{gainza2020deciphering, EsquivelRodriguez2014, Milanetti2021}. Intuitively, the more the surface is sampled the more the reaching for hot-spot is accurate. Similarly, the higher the number of different points used to represent the surface the higher the level of detail of the molecular shape. 

However, time and computational costs limit both the resolution of the surface and the number of patches that can be sampled, especially for large protein complexes and/or in analyses that involve a big set of surfaces, like for example molecular dynamics data.

Here, we present a new method for reducing the molecular surface, maximizing the overall information of the protein shape, and minimizing, in principle, the number of total surface points. 
The basic idea of the proposed new algorithm is the selection of molecular surface points according to the local roughness: increasing the sampling in high roughness and decreasing the sampling in the more flat regions. 
In particular, we define a sampling probability that depends on the local roughness of the surface. We discuss the performance of the algorithm as a function of several parameters, evaluating the ability of the algorithm to describe locally portions of the surface in comparison to uniform random sampling. 
To evaluate the resulting representation of the surface, we compared the shape similarity between a portion of the surface obtained via our algorithm and with uniform sampling.

Shape similarity is measured using a method we have recently developed, which is completely parameter-free and independent of the orientations of interacting structures \cite{Milanetti2021}. In fact, this is based on the description of each portion of the molecular surface by expanding the well-exposed molecular surface patches in terms of 2D \textit{Zernike} polynomials, in order to be able to measure the geometrical complementarity between interacting proteins, with lower computational cost than its 3D version \cite{Daberdaku2018,Kihara2011,Zhu2014,Venkatraman2009,DiRienzo2017,DiRienzo2020}.\\
We show that our proposed sampling reduces the number of considered points, minimizing the loss of information about the protein surface shape.

\section{Model and results}
The proposed method consists of a sampling suitably designed for selecting with greater probability surface points belonging to regions of high roughness.
Indeed, when choosing where to select the patches, a compromise between a too sparse sampling (which would lose some regions of the surface) and a too dense one (which would add no information and would instead increase the computational time) must be found.\\
To this end, we define a function used for the sampling of the original surface and we use a method based on the 2D formalism of the \textit{Zernike} polynomials to evaluate the ability of the algorithm to correctly approximate local regions of molecular surfaces with respect to uniform random sampling.
In the following, we describe in detail the designed algorithm, the method used based on the 2D \textit{Zernike} formalism, and the results obtained by varying the parameters.

\subsection{New roughness-dependent sampling}
To begin with, we numerically represent the molecular surface with a set of N points in the 3D space (the discretization of the surface). For each point $i$, we evaluated the exiting normal vector, $\bar{v}_i$, to the surface, originating from $i$. Next, we evaluate the local roughness of the molecular surface by looking at the relative orientation of the normal vectors with respect to each point $i$. 

\begin{figure}[H]
\centering
\includegraphics[width=0.48\textwidth]{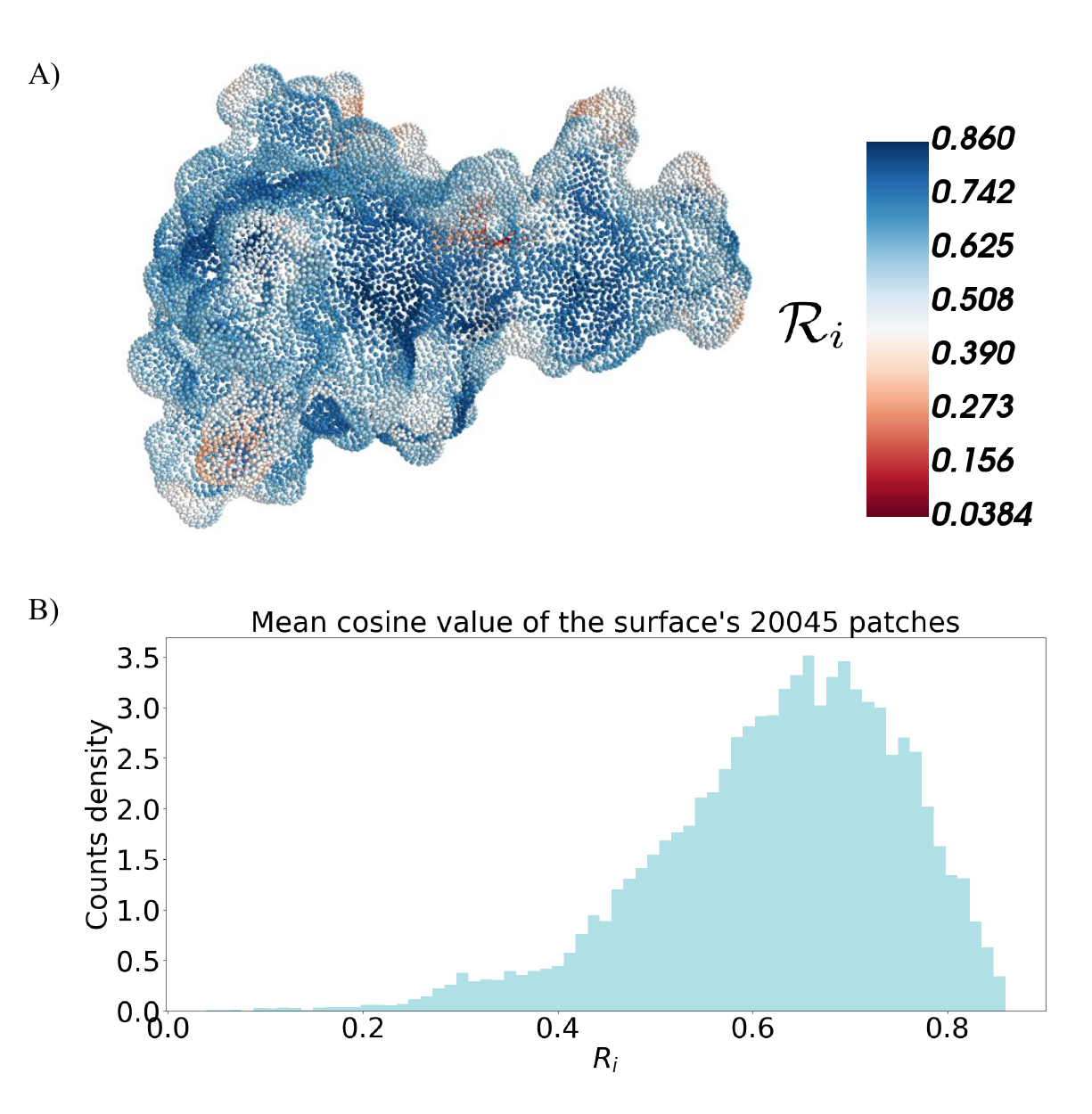}
\caption{A) Discretized representation of the molecular surface of the TDP-43 fragment 209-269 (PDB id: 4BS2). Each point of the surface is coloured according to the local roughness value, $\mathcal{R}_i$.\\ B) Distribution of the roughness, $\mathcal{R}_i$ found for each point $i$ of the considered surface.}
\label{cosine}
\end{figure}

To do so,  we define, starting from each point $i$, a patch with a sphere of radius $R_{patch}$ and we calculate the roughness as the mean of the cosines of the angles formed by the normal vectors associated to the $n_p$ points of the patch and the average normal vector:

\begin{equation}
    \mathcal{R}_i = \frac{1}{n_p}\sum_{j=1}^{n_p} \cos(\theta_{ij}) 
\end{equation}

with $\cos(\theta_{ij}) = \frac{ \bar{v}_i\cdot \bar{v}_j}{|\bar{v}_i| |\bar{v}_j|}$ and $\bar{v}_i = \frac{1}{n_p}\sum_j \bar{v}_j $. 
Figure \ref{cosine}a shows the molecular surface for a case-of-study protein, TDP-43 (PDB id: 4BS2) residues 209-296, colored according to the local roughness. Being a mean of cosines, the roughness ranges from zero to one (see Figure~\ref{cosine}b). When the considered patch is plane, the mean value of the cosine between the normal vectors of each point $i$ of the surface and the mean normal vector of that patch, $\mathcal{R}_i$, is close to one, while lower values of $\mathcal{R}_i$ indicate patches rougher.

Then, we associated to each point $j$ in the patch centered on a point $i$ the probability to be accepted for the sampling, defined as:
\begin{equation}\label{p}
p(j) = \alpha (1-\mathcal{R}_i)^\beta \left(\frac{r_{i,j}}{R_{patch}}\right)^{\gamma (1+\mathcal{R}_i)+\delta},
\end{equation}
where $r_{i,j}$ is the distance of the point $j$ of the patch from its center $i$, and $\alpha$, $\beta$, $\gamma$ and $\delta$ are parameters that can be optimized to yield different sampling scenarios.\\

For instance, when $\beta=0$, $\gamma=0$, and $\delta=0$, we obtain a uniform random extraction, whose total number of points depends on $\alpha$: this parameter determines the fraction (over all the surface's points) of considered points.

In general, when a patch $i$ has a high roughness, more points are needed to describe it. On the other hand when it is more plane we need fewer points, and indeed $(1-\mathcal{R}_i)$ becomes smaller. Finally, the center of a patch is always selected, but then to capture the surface's irregularities we can use as centers for the \textit{Zernike} patches the points further away from it, i.e. the ones with a high value of $r_{i,j}$. By elevating this term to the $(1+\mathcal{R}_i)$ we are changing the distribution of sampled points in each patch as a function of that patch roughness.

\subsection{Optimization of the sampling parameters}
Here, we describe the method used to determine how effectively, starting from a sampling (determined by the combination of the parameters $\alpha$, $\beta$, $\gamma$, and $\delta$), we can describe the original surface.\\
\begin{enumerate}
    \item For each combination of these four parameters, we sample from the original total surface a number $n_S$ of points and we define a new surface determined by these $n_S$ points. Then, we extract from the original total surface again $n_S$ points, but this time with a uniform distribution (or random extraction).
    \item We select from the total surface $n_{test}$ points, and define around each of them a region with radius $R=6~$\AA.
    \item Next, we associate to each one of these points, $j$, three vectors: $z_{tot}(j)$, $z_S(j)$ and $z_R(j)$. $z_{tot}(j)$ contains the \textit{Zernike} descriptors which describe that patch as defined by all the total points included in it, $z_S(j)$ describes the patch as defined by the sampled points included in it and $z_R(j)$ describes the patch as defined by the included randomly extracted points.
    \item For each of the $n_{test}$ patches we compute the distances $Z_{t-S}(j)=z_{tot}(j)-z_S(j)$ and $Z_{t-R}(j)=z_{tot}(j)-z_R(J)$. We average all the obtained $Z_{t-S}(j)$ and $Z_{t-R}(j)$, and obtain respectively the values $Z_{t-S}$ and $Z_{t-R}$. Since we are considering the description given by all the original points as our "ideal", for a good sampling we expect the value of $Z_{t-S}$ to be small, and in particular smaller than $Z_{t-R}$.
    \item Finally, we compute the difference $d=Z_{t-R}-Z_{t-S}$. The best sampling for a surface should result in the maximization of $d$.
\end{enumerate}
Our main objective is to describe the original surface as precisely as possible with a minimum number of points.\\
But to be more user-friendly we also add the option of pre-selecting a maximum number of sampled points and search for the best parameters combination with this constriction on $n_S$.\\

\subsection{Optimization starting from a pre-determined $n_S$}
\begin{figure*}[t]
\includegraphics[width=.8\textwidth]{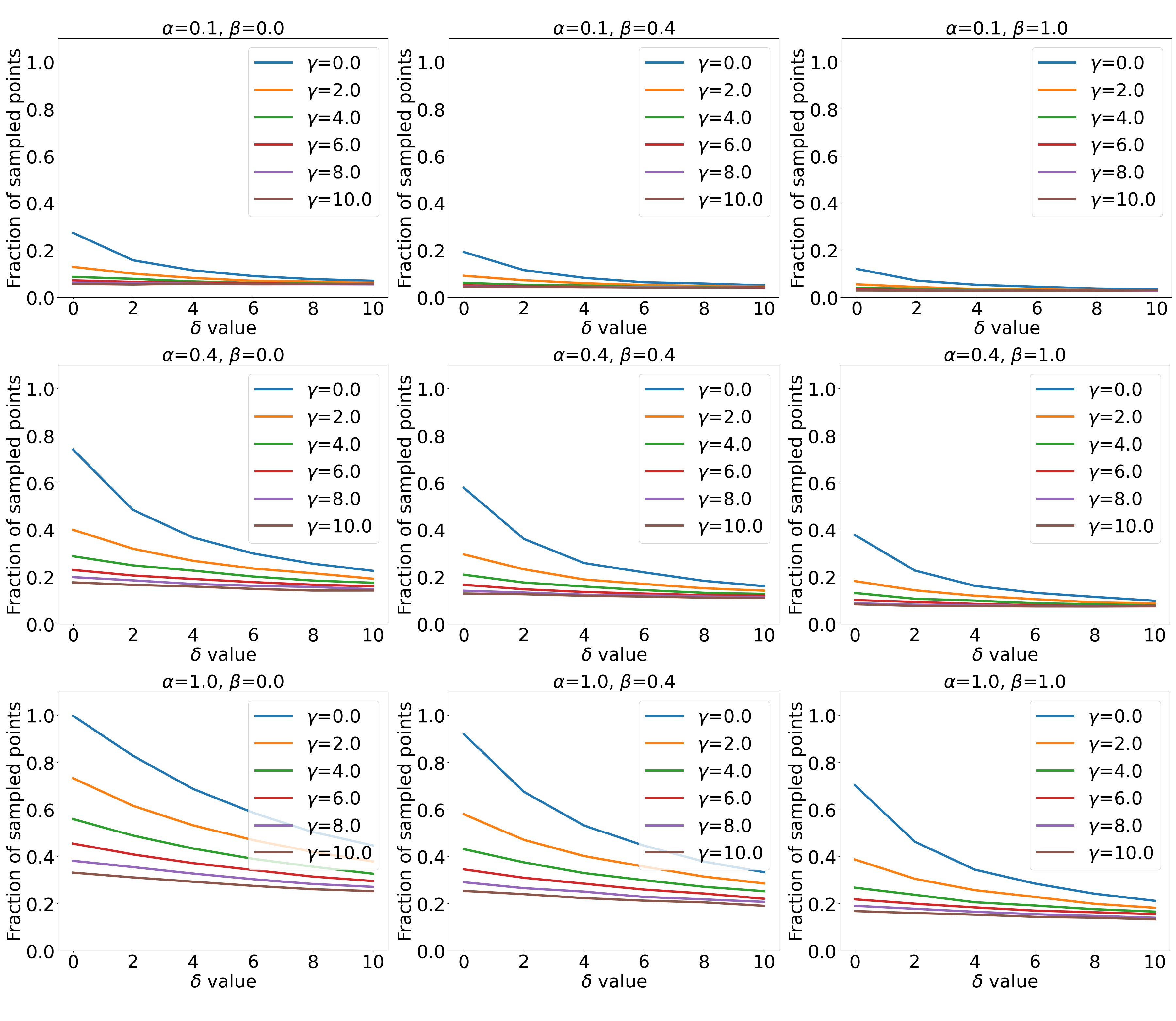}
\centering
\caption{Percentage of surface points that are sampled by varying $\alpha$, $\beta$, $\gamma$ and $\delta$. These plots were obtained after the sampling of the surface of A1.}
\label{n points}
\end{figure*}

Indeed, by varying the four parameters we can study a wide range of $n_S$. Figure \ref{n points} shows an example of how this number changes for changing combinations of parameters.

Table \ref{n_s} shows, for each arbitrarily determined range of $n_S$, the combination of parameters that results in the sampling with the highest $d$. 

\begin{table}[t]
\begin{center}
\begin{tabular}{ |c||c|c|c|c|c|c|  }
 \hline
Range of the sampled points & $\alpha$ & $\beta$ & $\gamma$ & $\delta$ & $d$ & $Z_{t-S}$ \\ 
 \hline 
\cellcolor{lightgray}{2.63\% - 12.34\%} & 0.4  & 0.2  &10.0  &10.0  &1.61  &7.36 \\
 \hline
\cellcolor{lightgray}{12.34\% - 22.05\%} & 1.0  & 0.4  &10.0  &6.0  &1.98  & 6.39 \\
\hline
\cellcolor{lightgray}{22.05\% - 31.76\%} & 1.0  & 0.2  &6.0  &10.0  & 2.31  & 5.85 \\
\hline
\cellcolor{lightgray}{31.76\% - 41.47\%} & 1.0  & 0.0  &6.0  & 4.0  & 2.46  & 4.97  \\
\hline
\cellcolor{lightgray}{41.47\% - 51.18\%} & 1.0  & 0.0  &6.0  & 0.0  & 2.69  & 4.10  \\
\hline
\cellcolor{lightgray}{51.18\% - 60.89\%} & 1.0  & 0.0  & 2.0  & 4.0  & 2.24  & 3.24  \\
\hline
\cellcolor{lightgray}{60.89\% - 70.60\%} & 1.0  & 0.2  & 0.0  & 4.0  & 1.70  & 2.53  \\
\hline
\cellcolor{lightgray}{70.60\% - 80.30\%} & 1.0  & 0.0  & 2.0  & 0.0  & 0.18  & 1.38  \\
\hline
\cellcolor{lightgray}{80.30\% - 90.01\%} & 1.0  & 0.0  & 0.0  & 2.0  & -0.14 & 0.81 \\
\hline
\cellcolor{lightgray}{90.01\% - 99.73\%} & 1.0  & 0.2  & 0.0  & 0.0  & -0.01 & 0.27   \\
\hline
\end{tabular} 
\caption{Best sampling for each range of selected points ($n_S$). }
\label{n_s}
\end{center} 
\end{table}

\subsection{Optimization without restriction }
\begin{figure}[t]
\includegraphics[width=\columnwidth]{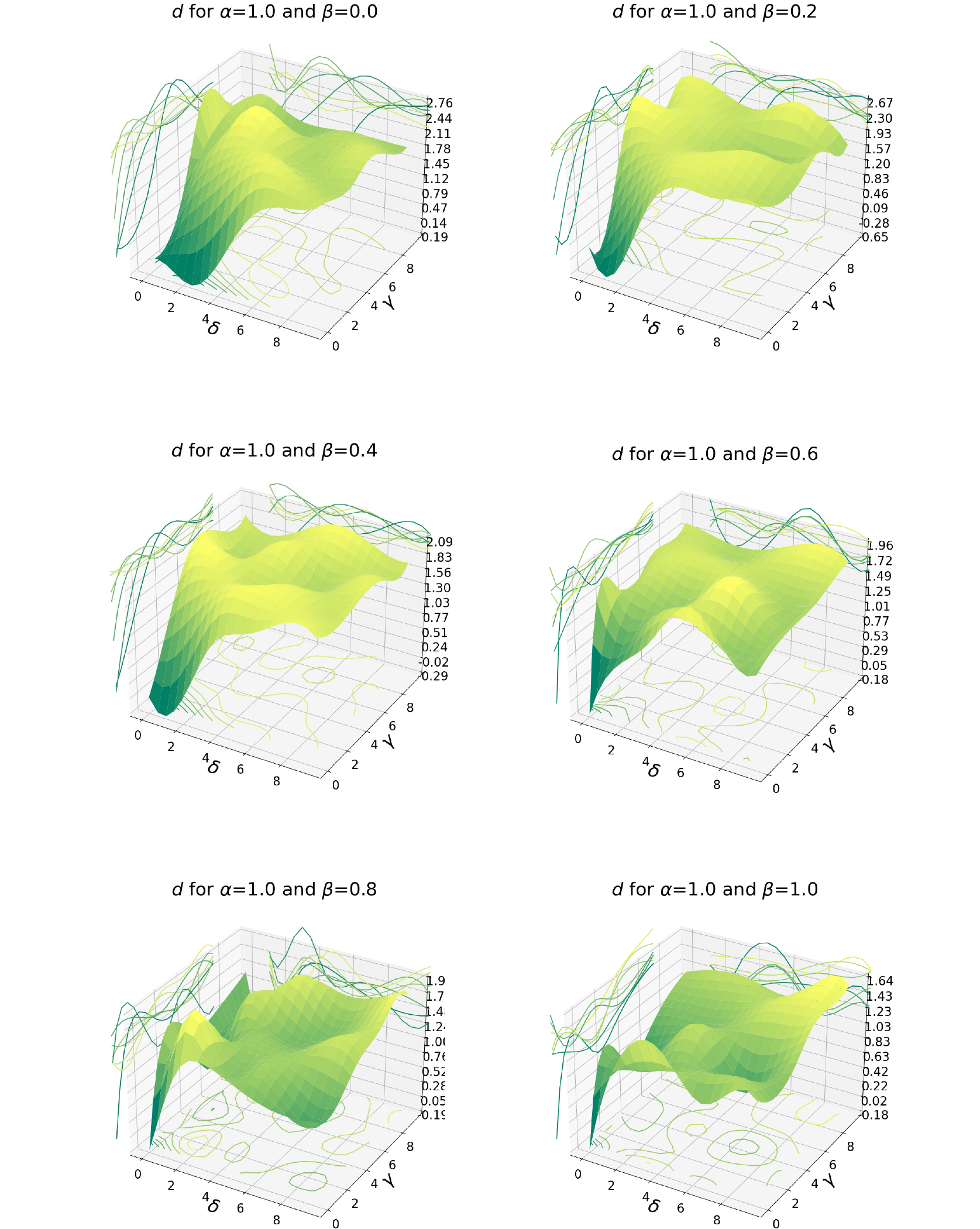}
\centering
\caption{$d$ as a function of varying values of $\gamma$ and $\delta$. Each plot is obtained with a fixed value for $\beta$, and for all the plots the same fixed value of $\alpha=1.0$ is used. The plotted surfaces are obtained with an interpolation of the effectively computed values, that were obtained with all the possible combination, for each $\beta$ value, of $\gamma=[0,2,4,6,8,10]$ and $\delta=[0,2,4,6,8,10]$. }
\label{d}
\end{figure}

When there is no restriction on the number of sampled points, we can fix $\alpha=1$, since it is a multiplicative parameter that causes no variation of the distribution of sampled points between patches with different roughness values. When $\alpha<1$ we are removing from each patch the same percentage of points, so if we have no requirements about the final $n_S$, this arbitrary (meaning that it does not depend on the surface shape) removal of points can be avoided.\\
Consequently, we are interested in finding the combination of $\beta,~\gamma$, and $\delta$ that results in the highest $d$. While it is true that a good sampling should result in a high $d$ combined with a low $n_S$, the weights that these two components should have in an optimization function will change according to the application and cannot be generalized.\\
Figure \ref{d} shows how $d$ changes as a function of $\delta$ and $\gamma$, for different fixed values of $\beta$.

\begin{figure*}[t]
\includegraphics[width=.8\textwidth]{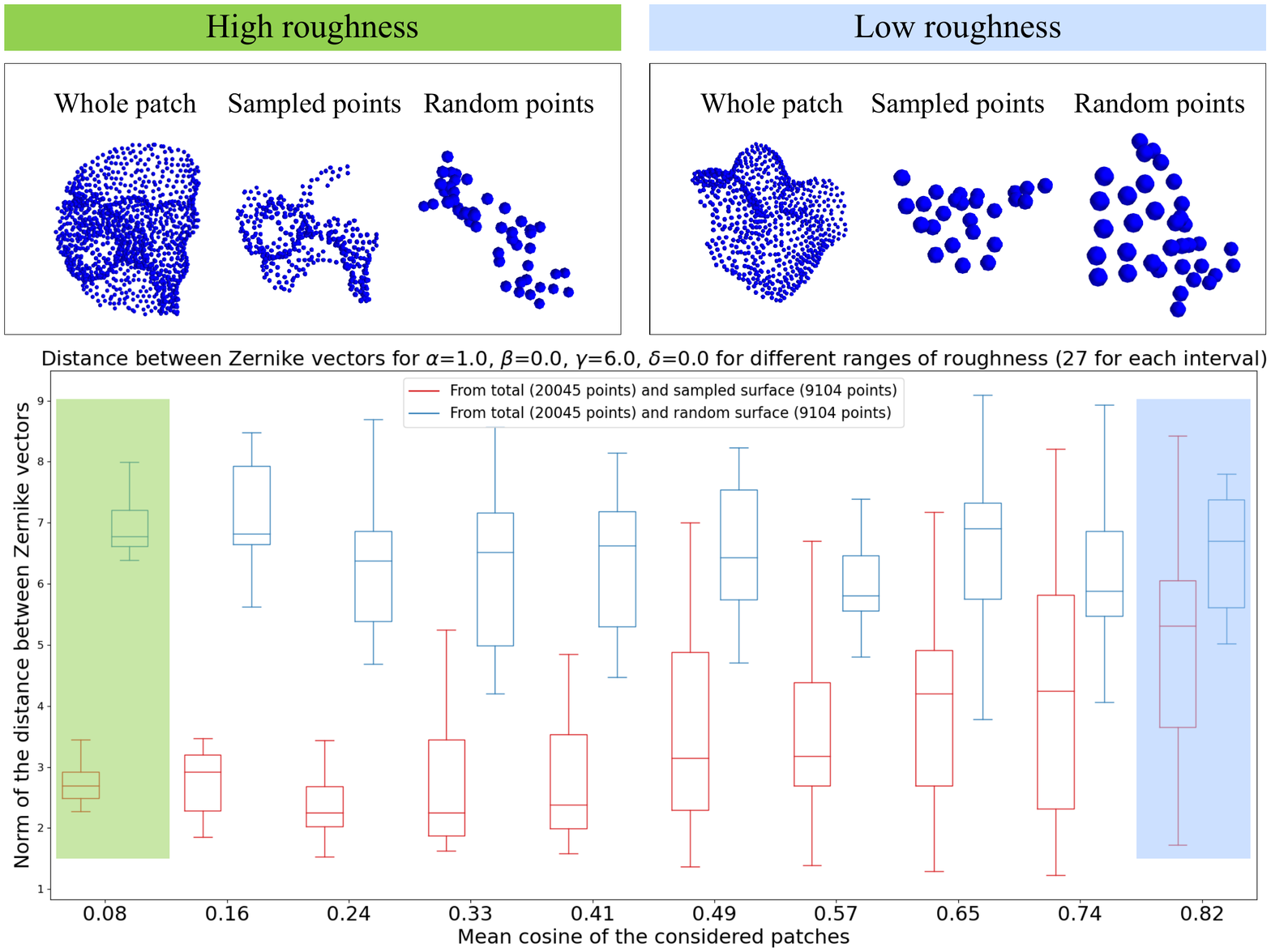}
\centering
\caption{Top: on the left, an example of how a rough patch is represented with all the points in the surface (\textit{whole patch}), with only the points resulting from the sampling (\textit{sampled points}) and with randomly extracted points (\textit{random points}). On the right, the same three representation cases for a plane patch.\\
Bottom: in red, the box plot for different ranges of patches' roughness of the $Z_{t-S}(j)$. In blue, the box plot for different ranges of patches' roughness of the $Z_{t-R}(j)$. This is in the case of a sampling with parameters $\alpha=1,~\beta=0,~\gamma=6$ and $\delta=0$ , whose combination results in the highest value of $d$ for the A1 surface.}
\label{vs}
\end{figure*}

\begin{figure}[t]
\includegraphics[width=.48\textwidth]{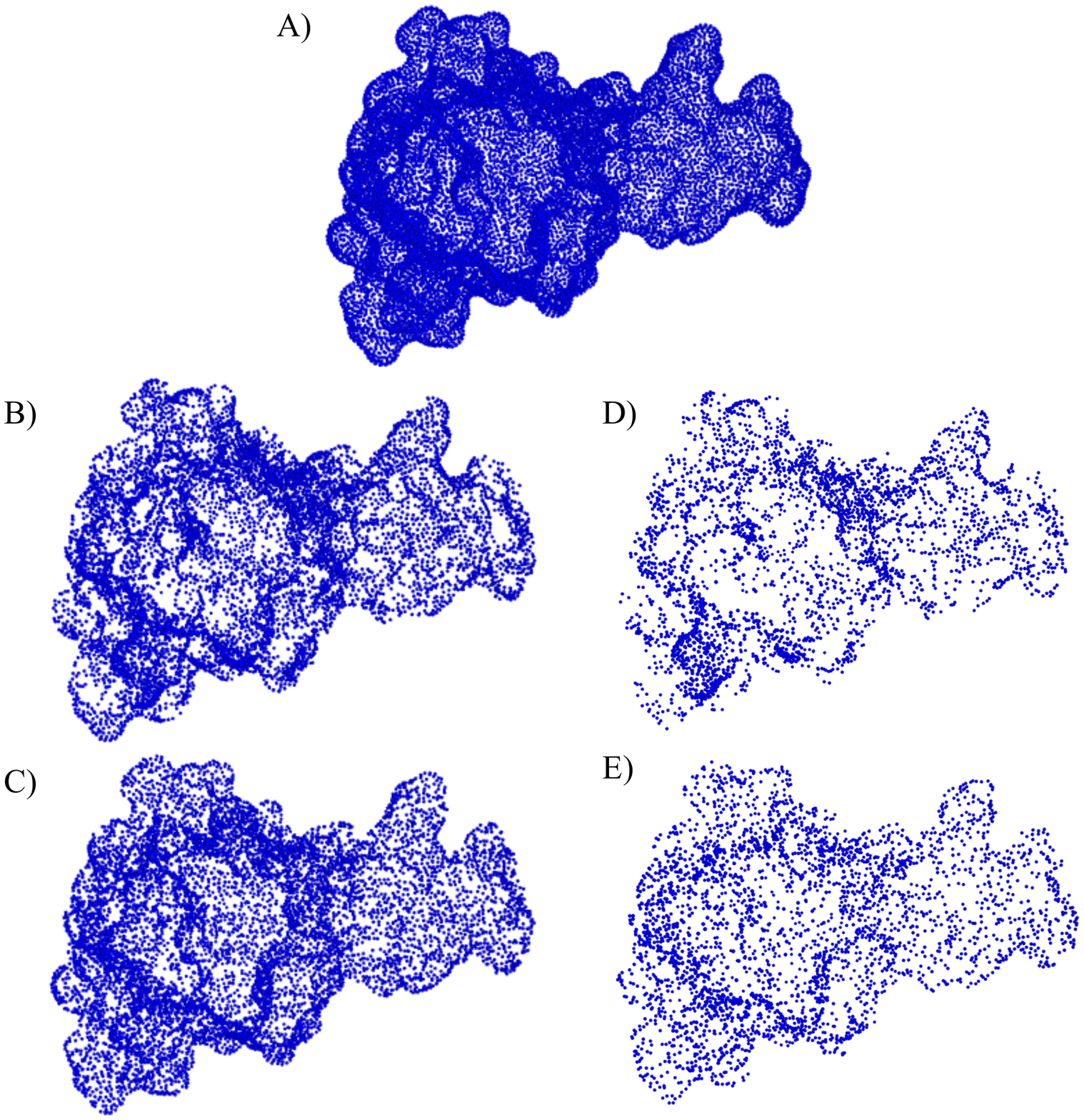}
\centering
\caption{A) 3D reconstruction of the A1 surface from all its surface points.\\
B) 3D reconstruction of the same surface, with a subset of the original points selected with the sampling with parameters $\alpha=1,~\beta=0,~\gamma=6$ and $\delta=0$. This is the best sampling for this surface.\\
C) 3D reconstruction of the same surface, with a subset of the original points. This subset counts the same number of points selected with the best sampling, but in this case, they were randomly extracted.\\
D) 3D reconstruction of the same surface with a sampling in which $\alpha=1,~\beta=1,~\gamma=10$ and $\delta=0$, chosen only because it results in fewer points than the best sampling of plot B) so that is easier to see by eye how the sampled points are gathered in the rough regions.\\
E) 3D reconstruction of the same surface, with the same number of points used in plot D) but randomly extracted.}
\label{screening}
\end{figure}

Once the best parameters combination has been detected, we can better understand what exactly makes the sampling better than a random extraction from the total surface of the same number of points. To understand the action of a sampling on a surface we have to look at what happens for different roughness of the patches, given its dependency in Equation \ref{p}. Figure \ref{vs} shows for example the box plot of $Z_{t-S}(j)$ and $Z_{t-S}(j)$ divided according to the roughness of the patches they describe. The considered patches are extracted (according to their roughness) from the points obtained with the best sampling of the considered surface.
As expected, we can see that plane patches (whose $\mathcal{R}_i$ tends to one) are reconstructed nearly equally with the sampling and the random extraction, whereas for increasing roughness ($\mathcal{R}_i$ approaching zero) the sampling results in \textit{Zernike} vectors much closer to the ones obtained from the total surface, compared to the random case. 
That said, it must be remembered that with the sampling we are concentrating the $n_S$ points on the roughest zones. This means that with the sampling the plane patches are described with fewer points than in the random case: the improvement for the rough patches comes from a better surface reconstruction, while for plane patches it derives from the fact that a smaller number of points is used to reach the same reconstruction.
\\[14pt]
Finally, Figure \ref{screening} shows an example of how the surface is described when all its points are considered versus when only some subsets are selected, with the sampling or randomly.

\section{Conclusions}
In this paper, we present a new sampling method to reduce the number of points needed to describe a molecular surface.
Given the importance of the definition of the molecular surface for the study of the interactions between molecules, we design a new algorithm to efficiently select the points of the original surface that minimize the loss of information in terms of describing the local shape of surface regions, which can potentially interact with a molecular partner. 
To test its performance, we use a recently developed method based on 2D \textit{Zernike} polynomials, capable of describing the shape properties of portions of molecular surfaces. 
By means of \textit{Zernike}, we verified if the patches centered around the sampled points were indeed the most representative of the surface. 
The sampling of these points is determined by Equation \ref{p} and, when the best parameters are selected, results in a satisfactory reconstruction of the surface that needs only a subset of the total surface's points.\\
To determine how accurate the method is in selecting a subset of surface points, we compare the results of the sampling with what is obtained with a random selection of the same number of points. This is done in terms of the mean distance between the \textit{Zernike} vectors that describe a patch containing all the surface points and the ones that describe the same region, but containing this time only the points selected with the sampling or randomly.\\
Up to now, we tested this new method on a limited set of surfaces (corresponding to the equilibrium configuration of two fragments of the RRM2 domain of the protein TDP-43, given that the reduced dimensionality of the system allowed us to carry out tests on the variation of the parameters with relatively low computational cost), but, in principle, the method is applicable to any protein. 
Once the best combination of parameters is associated with each kind of surface (depending on its mean roughness), this method will allow each user to immediately identify, given (if desired) the required number of sampled points and the mean roughness of the surfaces to be analyzed. Thanks to the reduction of the points that have to be considered, the computational cost of the following analysis on the surfaces is reduced.
Therefore, the method can be applied to (i) macromolecules composed of a high number of residues, (ii) the analysis of the molecular surface of frames obtained from molecular dynamics simulations, and (iii) molecular surfaces calculated at high resolution for which a high number of points is required for a more detailed description.

\section{Materials and Methods}

\subsection{Protein structures and computation of the molecular surfaces}   
The surface taken as an example for the application of this new sampling for the \textit{Zernike} method was obtained from a Molecular Dynamic (MD) simulation of the residues 209-269 of the protein TDP-43.\\
The starting point of this simulation was the selection of this range of residues from the Nuclear Magnetic Resonance (NMR) structure of the TDP-43 tandem RRMs in complex with UG-rich RNA (PDB id: 4BS2), available at the Protein Data Bank \cite{BERNSTEIN1977}.
The MD simulation was carried on for 10 $\mu s$, and all its steps were performed using Gromacs 2019.3~\cite{https://doi.org/10.5281/zenodo.3562495}.\\
The topologies of the system were built using the CHARMM-27 force field~\cite{charmm}, the standard force field for proteins.
The fragment was placed in a rhombic dodecahedron simulative box, with periodic boundary conditions, filled with 4607 TIP3P water molecules~\cite{Jorgensen1983}. 
The rhombic dodecahedron box is built so that each atom is at least at a distance of 11 \AA$~$from the box borders. Its volume is 71\% of the one of a cubic box of the same periodic distance: fewer water molecules have to be added to solvate the protein. For a protein to have the correct behavior there need to be at least two or three layers of water around it: with 11 \AA$~$there is space for approximately five layers.
The final system, consisting of 14777 atoms, was minimized with 102 steps. Each step had a size of $0.01$, while the force limit value was set to $max(|\textbf{F}_n|)<10^3~kJ/mol/nm$.\\
The thermalization and pressurization of the system in NVT and NPT environments were run each for 0.1 $ns$ at 2 $fs$ time-step. The temperature was kept constant at 300 $K$ with a Modified Berendsen thermostat and the final pressure was fixed at 1 $bar$ with the Parrinello-Rahman algorithm \cite{parrinello} (with a time constant of coupling between the system and the barostat of $\tau_P=2~ps$), which guarantees a water density close to the experimental value of the SPC/E model of water of $1008~kg/m^3$ ($1006\pm 5~kg/m^3$). LINCS algorithm \cite{lincs} was used to constraint h-bonds.\\
Finally, the system was simulated with a 2 $fs$ time-step for 10 $\mu s$ in periodic boundary conditions, using a cut-off of 12 \AA$~$ for the evaluation of short-range non-bonded interactions and the Particle Mesh Ewald method \cite{Cheatham1995} for the long-range electrostatic interactions.\\
For all these steps the Leap-Frog integrator and the Verlet cut-off scheme were used.
\\[10pt]
The aim of the simulation was to find the equilibrium configurations of this TDP-43 fragment: with this objective, we performed clustering of the PCA of the trajectory resulting from the MD simulation, and associate each centroid with an equilibrium configuration. Then, we extracted the PDB file describing one of these configurations: this is what we called A1. To compute for this structure the solvent-accessible surface, we used DMS \cite{richards1977areas}, with a density of 5 points per \AA$^2$ and a water probe radius of 1.4 \AA. For each surface point, we calculated the unit normal vector with the flag $-n$.\\

\subsection{Zernike 2D procedure}
The molecular surface we obtain with DMS is represented by a set of points in the three-dimensional space.
The first step of the 2D \textit{Zernike} algorithm is to select from the surface a patch $\Sigma$, defined as the set of surface points included in a spherical region having radius $R_{zernike}=6$ \AA$~$ and centered in one point of the
surface. The points not directly connected to that region of the surface (for example coming from a protuberance included in the sphere) are removed.
Once the patch has been selected, the mean vector of the normal vectors of the patch points is computed and oriented along the $z$-axis. Thus, given a point, $C$ on the $z$-axis, the angle $\theta$ is defined as the largest angle between the $z$-axis and a
secant connecting $C$ to any point of the surface $\Sigma$. $C$ is then set so that $\theta=45^\circ$ and each surface point is labeled with its distance $r$ from $C$. As a next step, a square grid that associates each pixel with the mean $r$ value calculated on the points inside it is built.
The gap of pixels where no point of the surface has been projected are filled by using the average value of the surrounding pixels.\\
Such a 2D function can now be expanded on the basis of the \textit{Zernike} polynomials.\\
Indeed, each function of two variables $f(r,\psi)$ defined in polar coordinates inside the region of the unitary circle ($r<1$) can be decomposed in the \textit{Zernike} basis as
\begin{equation}
f(r,\psi)=\sum_{n'=0}^\infty\sum_{m=0}^{n'}c_{n'm}Z_{n'm}(r,\psi),
\end{equation}
with
\begin{equation}
    c_{n'm}=\frac{n'+1}{\pi}\int_0^1dr~r\int_0^{2\pi}d\psi Z_{n'm}^*(r,\psi)f(r,\psi)
\end{equation}
and 
\begin{equation}
    Z_{n'm}=R_{n'm}(r)e^{im\psi}.
\end{equation}
$c_{n'm}$ are the expansion coefficients, while the complex functions $Z_{n'm}(r,\psi)$ are the \textit{Zernike} polynomials. The radial part $R_{n'm}$ is given by
\begin{equation}
    R_{n'm}(r)=\sum_{k=0}^{\frac{n'-m}{2}}\frac{(-1)^k(n'-k)!}{k!\big(\frac{n'+m}{2}-k\big)!\big(\frac{n'-m}{2}-k\big)!}.
\end{equation}
Since for each couple of polynomials, it is true that
\begin{equation}
    Z_{n'm}|Z_{n''m'}=\frac{\pi}{n'+1}\delta_{n'n''}\delta_{mm'},
\end{equation}
the complete sets of polynomials forms a basis, and knowing the set of complex coefficients ${c_{n'm}}$ allows for a univocal reconstruction of the original patch.\\
The norm of each coefficient $z_{n'm}=|c_{n'm}|$ constitutes one of the \textit{Zernike} invariant descriptors.

\subsection{Complementarity evaluation}
Once a patch is represented in terms of its Zernike descriptors, the shape relation between that patch and another one can be simply measured as
the Euclidean distance between the invariant vectors.
The relative orientation of the patches before the projection in the unitary circle must be considered. In fact, if we search for similar regions we must compare patches that have the same orientation once projected in the 2D plane, i.e. the solvent-exposed part of the surface must be oriented in the same direction for both patches, for example as the positive z-axis. If instead, we want to assess the complementarity between two patches, we must orient the patches contrariwise, i.e. one patch with the solvent-exposed part toward the positive z-axis (`up') and the other toward the negative z-axis (`down').

\end{document}